# Ontology-Based Administration of Web Directories

Marko Horvat, Gordan Gledec and Nikola Bogunović

Faculty of Electrical Engineering and Computing, University of Zagreb
Unska 3, HR-10000 Zagreb, Croatia
E-mail: {Marko.Horvat2, Gordan.Gledec, Nikola.Bogunovic}@fer.hr

**Abstract.** Administration of a Web directory and maintenance of its content and the associated structure is a delicate and labor intensive task performed exclusively by human domain experts. Subsequently there is an imminent risk of a directory structures becoming unbalanced, uneven and difficult to use to all except for a few users proficient with the particular Web directory and its domain. These problems emphasize the need to establish two important issues: *i)* generic and objective measures of Web directories structure quality, and *ii)* mechanism for fully automated development of a Web directory's structure. In this paper we demonstrate how to formally and fully integrate Web directories with the Semantic Web vision. We propose a set of criteria for evaluation of a Web directory's structure quality. Some criterion functions are based on heuristics while others require the application of ontologies. We also suggest an ontology-based algorithm for construction of Web directories. By using ontologies to describe the semantics of Web resources and Web directories' categories it is possible to define algorithms that can build or rearrange the structure of a Web directory. Assessment procedures can provide feedback and help steer the ontology-based construction process. The issues raised in the article can be equally applied to new and existing Web directories.

**Keywords:** Ontology, Ontology Alignment, Artificial Intelligence, Semantic Web, Web directory

## 1    Introduction

The Semantic Web vision and related spectrum of technologies have enjoyed rapid development during the last ten years. The initial and often cited paper by Tim Berners-Lee [1] introduced a rather abstract notion of universally described semantics of information and services on the Web. The vision of a Web as a shared common medium for data, information and knowledge exchange, and collaboration, fostered a wealth of research and pragmatic development. The idea itself was simple but appropriately far reaching. The Semantic Web brought the power of managed expressivity provided by ontologies to the World Wide Web (WWW) [2]. Today research in Semantic Web applications is very diverse but not particularly focused on the problem of ontologically-based Web directories. So far only a handful or papers have been published on the topic of combining ontologies and Web directories [3][4][5]. Furthermore, as yet a lot of the effort is unfinished and more computer systems utilizing ontologies are in the phase of research and development (R&D) than in everyday production [6].

However, Web directories have simple hierarchical structures which are effective in data storage and classification. This makes Web directories important applications for storing information and its taxonomy, but also motivates research in the assessment of their semantic qualities and automatic management of their hierarchical structures. Solutions to both problems can also be useful in the more general and commonplace problem of ontology sorting.

It should be mentioned that Web directories are important but often overlooked means of resource integration and implementation of collective intelligence on the Web. The construction and maintenance of Web directories are both asymmetrical and collectively executed tasks where the contributors provide resources (i.e. information) to the directory and its administrators decide if the resources can be accepted and where should they be placed within the directory's structure. The third party in these processes, the general users, can extract semantically ordered data from the directories and freely use them in their own business processes. It can be said that Web directories are public frameworks for information sharing and collaboration. As such they are designed for hierarchical data storage and retrieval by means of browsing. Furthermore, in the context of the Semantic Web vision, Web directories are taxonomies of semantically and formally annotated data. The information stored in the directories is formatted in machine and human readable form, and thus becomes extractable by intelligent agents and can be used in distributed intelligent systems.

The remainder of the paper is organized as follows; the next section describes the categories and the structure of Web directories. Mutual associations between the Semantic Web and Web directories, as well as the semantic dimension of categories, are all presented in the third section. Web directories construction scenarios are presented in the fourth section, while the fifth chapter describes an algorithm for their ontology-based construction. Semantic quality measures of Web directories are discussed in the sixth section. Related publications and our conclusion with an outlook for future work are presented at the end of the paper.

## 2      Categories and the structure of Web directories

In order to explain how Web directories can be positioned within the Semantic Web vision, it is necessary to formally define a Web directory, its constituent components and their organization. It is also important to add semantic annotations to these building elements.

A web directory (web catalog or link directory, as it is also called) is a structured and hierarchically arranged collection of links to other web sites. Web directories are divided into categories and subcategories with a single top category, often called the root category, or just the root. Each category has a provisional number of subcategories with each subcategory further subsuming any number of other subcategories, and so on. Furthermore, every category has a unique name with an accompanying Uniform Resource Locator (URL), and can also carry other associated information.

Each category of a Web directory contains a set of links to various sites on the WWW and another set of links to other categories within the web directory. These two sets of links represent the most important characteristic of any Web directory. Links to categories within the same directory are called cross-links.

Each Web directory has to have a start page, i.e. a home page, which represents its root category, and every other category of a Web directory has its own adjoined web page. The start page displays subcategories that belong to the root. By following a link to a subcategory, user opens that category's page and browses through its links and subcategories. This process continues until the user finds a link to a web resource that s/he is looking for. In essence, the user can be described as an intelligent agent that traverses the structure of a Web directory looking for specific information.

Since Web directories are always rooted and the order of categories is strictly maintained, it is possible to assign level numbers to categories. The subcategories of the root are the 2$^{nd}$ level categories, and in turn their subcategories are the 3$^{rd}$ level categories, and so on. As a convention the root is always a 1$^{st}$ level category. The maximum level of a Web directory is called depth.

Each category, except the root, has one category above it, which is called its parent. The categories below a certain category (i.e. with a greater level number than the category) are called its children, while categories on the same level as a node are called its siblings. Categories with no children are called terminal categories, and a category with at least one child is sometimes called nonterminal category. Associations between categories are arbitrary, but there must be at least one path between any pair of categories. Disjoint sets of categories are not allowed, as well as parallel links and self-loops. Each nonterminal category must have links to all its children, but can also have links to other categories in the Web directory which are semantically similar, or otherwise analogous to the category.

## 2.1    Formal definition of Web directories

We shall formally designate with *C* the set of all categories in a Web directory, and *R* will be the set of all Web resources in a Web directory. One category with unique identification number *n* is denoted $c_n$. Category has its own characteristic URL *url*. The category $c_n$ must be a member of *C*. $C_n$ is a subset of *C* that belongs to the category $c_n$, and $R_n$ the subset of *R* with Web resources that belong to the category $c_n$. In order to be more informative, the categories can also be written as $c_n^l$ with their member level *l*, where *l* is a natural number smaller than or equal to the depth of a Web directory *L* [7]. The first level, or the root level, is always *l*=1. Category is a tuple $c_n = \{n, l, url, C_n, R_n\}$ and can be schematically annotated as in the figure below.

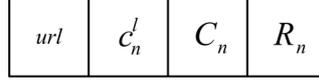

**Fig. 1.** Vector representation of a single category within a Web directory.

We can define a Web directory *wd* to be an element of the WWW. With **C** and **R** being members of *wd* the algebraic definitions of the elements of a Web directory and their mutual relationships are

$$\begin{aligned} C_n \in c_n \subset \mathbf{C} \\ R_n \in c_n \subset \mathbf{R} \\ l \in [1, L] \in \mathbf{N} \\ \mathbf{C} \in wd \subset \text{WWW} \\ \mathbf{R} \in wd \subset \text{WWW} \end{aligned} \quad (1)$$

The set of all children categories to $c_n^l$ is $C_n$. If the root is the top node (*l*=1) then the set of all children one level below is $C_n^{l+1}$ or (for the sake of brevity) $C_n^{+1}$, two levels below $C_n^{l+2}$ or $C_n^{+2}$, etc. As can be seen in (1), category is also a Web resource ($c_n \subset \mathbf{R}$), as it should be expected since it has unique URL and carries specific information. Furthermore, Web directory itself also becomes a tuple $wd = \{\mathbf{C}, \mathbf{R}\}$. The semantic information attached to resources and their related categories derives from the documents (e.g. Web pages, articles, blog posts, various textual documents, etc.) that are linked to the category. Also, the categories may have their own keywords and description defined by the directory's administrators. All this data collectively forms the category's semantic content.

Mathematically speaking, Web directories are simple rooted graphs [8]. In this formal respect, categories represent vertices and connections represent edges. The path between two vertices is called the arc, edge or link, and when there is an edge connecting the two vertices, we say that the vertices are adjacent to one another and that the edge is incident on both vertices. The degree of a vertex is equivalent to the number of edges incident on it.

Using the described formalisms, the schema of a simple Web directory with 6 categories distributed in 4 levels, with parent-child associations and two specific links $c_6 \rightarrow c_5$ and $c_3 \rightarrow c_2$ could be depicted with Fig. 2.

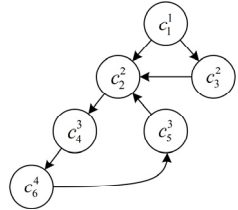

**Fig. 2.** Schematic representation of Web directory as a simple rooted graph.

However, the structure of a realistic Web category (Fig. 3) cannot be described just as a simple graph because of the cross-links which can define additional connections between categories. Apart from paths which connect categories with parent-child relationships, cross-links can associate any two categories.

Cross-links are added *ad hoc* by the Web directory's administrator staff to closely bind together two categories with similar semantic content. The nature of the category semantics and how they relate to one another has to be evaluated by the human administrator. Cross-links are very useful in facilitation of directory's browsing and information retrieval. They will allow users to find the needed Web resource faster in less steps and click-throughs. However, cross-links can, and most often do, form closed category loops. Excessive or unsystematic use of cross-links will make browsing more difficult and the process of information retrieval confusing for a user. In such circumstances users often will not follow the shortest browsing path through directory's categories, but rather will be sidetracked or deflected from their goal category. The semantic quality measures proposed in this paper address this very problem. They serve as objective criteria for the evaluation of Web directories structure in terms of its browsing convenience and overall usability.

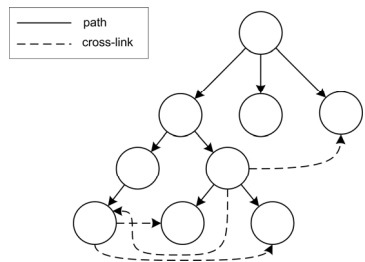

**Fig. 3.** Realistic Web directory with cross-links which allow loops and multiple paths between any two categories.

But if, for the sake of discussion, all categories of a Web directory except the root had paths only to its children, such structure would constitute a rooted tree as in Fig. 4.

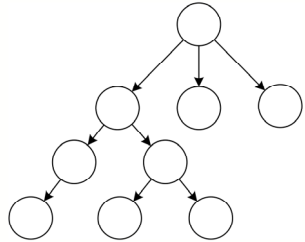

**Fig. 4.** Idealistic Web directory with only one path between any two categories.

In some cases the order of categories appearance may be relevant, e.g. the position of links within a category's Web page is prioritized, and in that case a Web directory can be formally modeled as an ordered and rooted simple graph.

Although the categorization of a Web directory should be defined by a standard and unchanging policy, this is frequently not the case. Web directories often allow site owners to directly submit their site for inclusion, even suggest an appropriate category for the site, and have administrators review the submissions. The directory's administrators must approve the submission and decide in which category to put the link in. However, rules that influence the editors' decision are not completely objective and are thus difficult to implement unambiguously. Sometimes a site will fall in two or even more categories, or require a new category.

Defining a new category is a very delicate task because it has to adequately represent a number of sites, avoid interfering with domains of other categories, and at the same time the width and depth of the entire directory's structure has to be balanced. A Web directory with elaborate structure at one end and sparse and shallow at the other is confusing for users and difficult to find quality information in.

Furthermore, after several sites have been added to a directory it may become apparent that an entirely new categorization could better represent the directory's content. In this case a part of directory's structure or even all of its levels have to be rearranged which is time consuming and labor intensive task. Reshuffling of the whole directory's structure may be warranted in the test phase, before the directory has entered the full production. But after users have grown accustomed to a certain structure, however suboptimal it may be, it would be unwise to profoundly alter the directory's shape. This would lead to renewed learning phase for all existing users and eventually may put them off in using the directory. Therefore, during exploitation the directory's structure should be altered only be adding and deleting cross-links, not the proper paths between parent-child categories. Alteration of cross-paths will not have such antagonistic effect towards the directory's users as the complete reassembly of directory's paths. But this only emphasizes the importance of creating the near-optimal structure during directory's construction which will have to be changed as little as possible later during the directory's lifetime.

## 3 The Semantic Web and Web directories

At the moment, resources available on the WWW are designed primarily for human and not machine use [9]. In order words, knowledge, declarative and procedural, offered by various Web sources is shaped in a way that better suites humans and not machines. The vision of the Semantic Web is directly aimed at solving this dichotomy by introducing self-describing documents that carry data and the accompanying metadata together, and thus organize and interconnect available information so it also becomes processable by computer applications [10].

The structure of a Web directory is basically a subjective construction. It depends on human comprehension and the policy taken by the Web directory's administrator, or even on the users that submit sites to the directory. It is important to note that not all Web directories, or even all segments of a Web directory, have the same editorial policy. Clearly, for the sake of a Web directory's informative clarity and usability, the semantic distance between any two categories should be approximately constant, and not dramatically vary from one category to the next. Whilst, the key for the selection of concepts that represent categories should remain uniform throughout the directory's structure. The only parameters that should be used to judge the quality of a directory are its informative value and usability, to humans and machines equally. In the fifth chapter we will propose several numerical parameters that objectively measure the worth of a directory.

### 3.1 A formal model of Web directories semantic content

In defining a formal model of Web directories semantic content it is necessary to assume that we have at a disposition function *sem* that takes a resource $r_i \in \boldsymbol{R}$ and from its semantic content builds an ontology $o_i \in \boldsymbol{O}$ where $\boldsymbol{R}$ and $\boldsymbol{O}$ are sets of all resources and ontologies, respectively.

$$sem : \boldsymbol{R} \rightarrow \boldsymbol{O} \qquad (2)$$

The function *sem* builds an ontology from a resource. In slightly different terms, it creates a solid representation of an abstract property. This property can be described as informal and explicit on the semantic continuum scale [11] and its technical realization is strictly formal. Operations of the function *sem* can be performed by a computer system or a domain expert, in which case we talk about automatic or manual ontology construction, respectively. The necessary mathematical assumption on *sem* is it has well-defined addition and subtraction operators in $\boldsymbol{R}$ and $\boldsymbol{O}$

$$\oplus : \boldsymbol{R} \times \boldsymbol{R} \rightarrow \boldsymbol{R} \qquad (3)$$

$$\odot : \boldsymbol{R} \times \boldsymbol{R} \rightarrow \boldsymbol{R}$$

$$\hat{+} : \boldsymbol{O} \times \boldsymbol{O} \rightarrow \boldsymbol{O}$$

$$\hat{-} : O \times O \to O$$

This allows application of union operator across these two sets and concatenation of individual resources and ontologies, as well as determining their respective differences

$$sem(r_1 \oplus r_2) = sem(r_1) \hat{+} sem(r_2) \quad (4)$$

$$sem(r_1 \odot r_2) = sem(r_1) \hat{-} sem(r_2)$$

We also define a modulo operator $|\bullet|$ on $O$ as

$$|\bullet| : O \times O \to O. \quad (5)$$

The *semantic content of a category* can be defined in three different ways:
1) by its Web resources
2) from subsumed categories
3) as a constant.

By the first definition, semantic content of a category $c_i$ within a Web directory *wd* derives from the semantic content of all its Web resources $r_{ij}$ where $r_{ij} \in R_i \in c_i$ as

$$sem(c_i) = \hat{\underset{r_{ij} \in R_i}{+}} sem(r_{ij}) \quad (6)$$

According to the second definition, the semantic content of $c_i$ can also equal the aggregation of the semantic content of its children categories $c_j \in C_i^{+1} \in c_i$

$$sem(c_i) = \hat{\underset{c_j \in C_i^{-1}}{+}} sem(c_j) \quad (7)$$

Finally, if $c_i$ has no resources $(R_i = \varnothing)$ and subcategories $(C_i = \varnothing)$ it is assumed that the semantic content of $c_i$ is defined by a constant $const_i$ as

$$sem(c_i) = const_i : R_i = \varnothing, C_i = \varnothing \quad (8)$$

The reasoning behind such threefold definition is that the meaning of categories is conformed to the directory's editorial policy. If a category is empty and no resources have been added, it will still have some member semantics attached by the Web directory administrator.

The structure of directory *wd* is *ideal* if for non-empty **R** and **C**

$$\left| \hat{\underset{r_{ij} \in R_i}{+}} sem(r_{ij}) \stackrel{\frown}{-} \hat{\underset{c_j \in C_i}{+}} sem(c_j) \right| = \varnothing \qquad (9)$$

$$\forall c_i \in wd, R_i \in c_i, C_i \in c_i$$

That is, the structure of directory *wd* can be considered perfect if and only if for each category $c_i \in wd$ the semantic content of its Web resources $R_i \in c_i$ and subsumed categories $C_i \in c_i$ are equal.

Pragmatically, we can define a neighborhood $\varepsilon$ within ***O*** and say that the structure of directory *wd* is *realistically ideal* if

$$\left| \hat{\underset{r_{ij} \in R_i}{+}} sem(r_{ij}) \stackrel{\frown}{-} \hat{\underset{c_j \in C_i}{+}} sem(c_j) \right| \leq \varepsilon \qquad (10)$$

$$\forall c_i \in wd, R_i \in c_i, C_i \in c_i$$

The existence of the function *sem*, with the described properties, is fundamental and indivertible in the ontology-based construction of Web directories.

## 4     Construction scenarios

The process of building Web directories has three actors [12]:

1. Web directory system (System)
2. Web directory administrator (Admin)
3. Administrator of a Web site listed in the Web directory (Site)

Ontology-based building process involves the same three actors and represents a subset of the general building process. This process includes three main tasks, or actions, that have to be performed by actors in order to construct a Web directory:

1. Semantics identification task (ID)
2. Semantics assignment task (ASSIGN)
3. Web directory addition task (ADD)

Semantics identification task is a process that recommends which ontology class, or classes, should be instantiated and assigned to a given Web resource. Semantics assignment task is a process that follows semantics identification, and actually assigns a set of ontology classes to a resource. Classes that are recommended and assigned do not necessarily have to be identical. If an actor has made an error and recommended the wrong class, the actor performing assignment can overrule his/her recommendation. Finally, when a set of classes has been assigned to a Web resource, it has to be added to a directory. Web directory addition task decides exactly where in a directory's structure the new resource will be placed. This is a complicated task because it can involve creation of an entirely new category, reshuffling and updating existing categories (both horizontally and vertically within the directory's structure), or simply adding the resource to an existing category. The order of these tasks and their mutual interaction is described in the following UML activity diagram (Fig. 3).

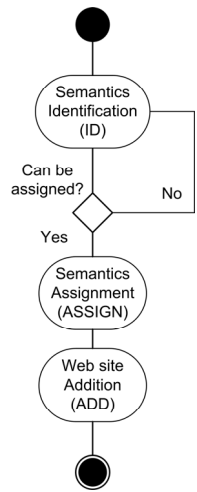

**Fig. 5.** Main tasks in ontology-based construction of Web directories.

Construction process scenarios can be divided in two groups:

1. Prevalently automated scenario (AUTO)
2. Mostly manual scenario (MANUAL)

Each scenario has several possible variations or sub-scenarios. Scenarios are distinguished by the level of human participation. Sub-scenarios describe the roles of the actors involved.

Utilization of human intelligence in majority of tasks is presumed in MANUAL scenario, while in AUTO scenario the Web directory computer system performs more tasks than human actors. In an ideal AUTO scenario the computer executes all tasks independently. Table 1 depicts all scenarios and their variations with respective grades of favorability. By following the highest grades in each scenario it is possible to determine the best actor for each task. Sequences of the best choices for each task are shown in UML diagrams in Fig. 6 and Fig. 7. Data in the table, temporally structured in the diagrams, reflects the "Best Practice" experience gathered during 15 years of administrating the Croatian Web directory [13].

**Table 1.** – Favorability of actors and task allocation in ontology-based construction of a Web directory.[1]

| Tasks | Roles | | |
|---|---|---|---|
| **MANUAL scenario** | Site | Admin | System |
| ID | ++ | + | n/a |
| ASSIGN | + | +++ | ++ |
| ADD | - | +++ | +++ |
| **AUTO scenario** | Site | Admin | System |
| ID | + | + | ++ |
| ASSIGN | - | +++ | ++ |
| ADD | - | + | +++ |

As can be seen in MANUAL scenario, Site is the best actor to perform ID, and Admin for ASSIGN. In this scenario ID is intentionally performed only by a human actor. ADD can be executed equally good by Admin or System, but it would be wrong to leave this task to Site. The reasoning behind allocation of actors in this scenario is that Site is the least dependable actor and its contribution is the most likely to be subjective and erroneous. The task will be most successfully performed by Admin, but it would be inefficient and wrong to give all tasks only to Admin. After all, one of the principal goals of the proposed system is to alleviate the burden of Web directory administration from the amenable personnel, and not to leave them with an equally difficult job. The best option is to allocate ID to Site and to leave the final decision about semantics to Admin who is the most knowledgeable and dependable actor of the three.

Much the same reasoning is reflected in the AUTO scenario; however the importance of System in this scenario is emphasized. Thus, System is the optimal choice for executing ID and ADD. Again, Admin will perform the final assignment of ontologies to resources (i.e. ASSIGN) to reduce possible errors to a minimum. In this scenario it was determined that it would be negative to let Site to execute ASSIGN and ADD since Admin or System can perform a better job at this tasks. In this scenario Site and Admin are equally suitable to execute ID. If ASSIGN is also given to System then the Web directory building system is fully automated.

---

[1] Sub-scenario grades: +++: the most acceptable, ++ favorable, + positive, – negative/unfavorable scenario, n/a not applicable.

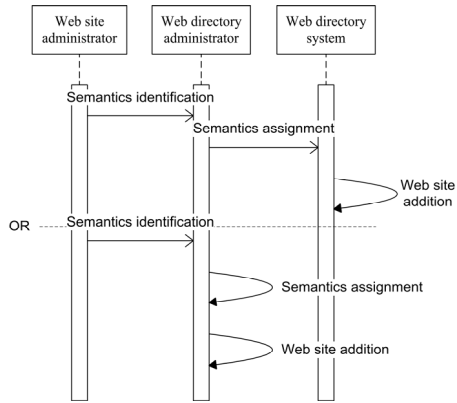 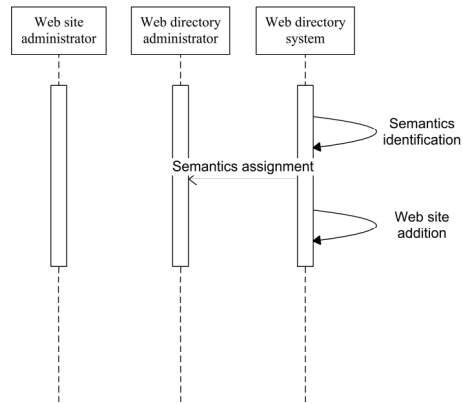

**Fig. 6.** UML sequence diagram with the selection of best actors in the MANUAL scenario.

**Fig. 7.** UML sequence diagram with the selection of best actors in the AUTO scenario.

## 5   Ontologically based construction

If it is possible to assign ontology to a Web resource and execute semantics identification and semantics assignment tasks as outlined in the previous chapter, it is also possible to define an ontology-based algorithm for automated construction of a Web directory structure. Such algorithm performs all tasks outlined in Fig. 2. The algorithm's input are links to Web resources that are being added to the Web directory, and output is schema of the directory. Schema can be represented in a number of ways, e.g. as a markup language, or additionally the algorithm can use the schema to automatically build the directory by writing and storing necessary static and dynamic Web files like HTML, JavaScript, PHP, etc.

In order to be able to define the described algorithm we will assume that we have at a disposition function *sem* as explained in 2.1 .

The basis for the algorithm construction process is the definition of category $c_i$ and its set of ontologies $O_i$ as a unified pair ($c_i$, $O_i$). In acquiring $O_i$ the algorithm uses the function *sem* and treats $c_i$ as a Web resource. The input is a set of Web resources **R** and the algorithm picks one resource $r_i$ at the time, translates in into an ontology $o_{NEW}$ and calculates the distance between $o_{NEW}$ and every ontology in the Web directory **O** looking for the closest. Categories are compared using their member ontologies. At each moment *wd* has *n* categories and a new category has index *n+1*.

Pseudocode for ontology-based construction of Web directories

```
// add the root category in web directory
add category (c₁, Ø);
// iterate through all web resources
for each rᵢ in R
{
 create new ontology instance O_NEW from resource rᵢ;
 if (K(C) = 1)
 {
  // if web directory contains only the root...
  create category (c_NEW, O_NEW);
  add rᵢ in c_NEW;
  add c_NEW in wd as C_{n+1}^{l+1};
 }
 else
 {
  // if web directory contains more categories...
  find the closest category (C_n^l, O_n) to O_NEW;
  d = dist(O_NEW, O_n);
  if (d > mindist_V)
  {
   create category (c_NEW, O_NEW);
   add rᵢ in c_NEW;
   add c_NEW in wd as C_{n+1}^{l+1};
  }
  else if (d > mindist_H)
  {
   create category (c_NEW, O_NEW);
   add rᵢ in c_NEW;
   add c_NEW in wd as C_{n+1}^{l};
  }
  else
  {
   add rᵢ in C_n^l;
  }
 }
}
```

The most significant aspect of the algorithm is reliance on ontologies and ontology aligning methods in order to measure similarity between ontologies and determine their mutual distance. The similarity measure $sim : \mathbf{C}^2 \rightarrow [0,1]$ between the two categories $c_1, c_2 \in \mathbf{C}$ and the distance function $dist(c_1, c_2) = 1 / sim(c_1, c_2)$ is defined in [14][15][16]. The algorithm uses two constants in a predefined metric; *minimal horizontal semantic distance* (*mindist_H*) and *minimal vertical semantic distance* (*mindist_V*) as thresholds in the category addition process. When a new category $c_j$ is being added and category $c_i$ already exists in $wd$ if $dist(c_i,c_j) > mindist_H$ then the algorithm will add $c_j$ as a new category of $wd$. Likewise, if $dist(c_i,c_j) > mindist_V$ then $c_j$ will be added in a new level of the directory $wd$, below $c_i$. If $dist(c_i,c_j) <= mindist_H$ AND $dist(c_i,c_j) <= mindist_V$ the algorithm will merge semantics of $c_j$ and $c_i$ incrementing initial ontology of $c_i$. Therefore, the thresholds are used in deciding whether it is necessary to add a new category in the directory's structure or to use an existing category. Also, the thresholds indicate where to add a new category: in the same level next to an existing category or below it.

The algorithm has two main branches. The first branch recognizes one special case when cardinal number K of all categories $C$ in $wd$ is 1, and the second branch processes three cases with cardinality of categories greater than 1. If $K(C) = 1$ then $l = 1$ and only the root category has been added to $wd$. In this case it is not necessary to calculate the distance between ontologies and a new category can be immediately constructed. If $K(C) > 1$ there are more categories, not just the root, and links to Web resources are assigned to the semantically closest categories. New categories are created if needed.

The single root node does not have a set of links to Web resources ($R_1 = \emptyset$) and it is assigned to an empty ontology ($c_1, \emptyset$), however the algorithm can be modified so it allows predefinition of main topics in a Web portal or Web directory according to the desired administrating policies.

The proposed algorithm is simple because it represents the direct and the most obvious implementation of an ontological principle in Web directory construction. Categories cannot be mutually prioritized, and the end structure is completely dependent on the order of links to Web resources which are the algorithm's input. Furthermore, there is no back-tracking or iterative optimization. For these reasons the algorithm may also be called basic or elementary, since all other ontology-based algorithms should provide better results. It could be used as an etalon for comparison of different algorithms for the construction of Web directories.

Execution of this algorithm can be assigned to different roles in MANUAL and AUTO scenarios (see Table 1). For example, a part of the algorithm, like ASSIGN can be given to human experts (System or Admin) and other tasks – ID and especially ADD – can be executed by an intelligent agent (System). Different assignment will yield diverse results and this presents an interesting topic for further study and experiments.

## 6    Semantic quality measures

During or after Web directory's construction it is highly desirable to establish some measures of value of the accomplished process. The criterion functions that will provide these measurements should be objective and universal. Benefits of such measures would be twofold: *i*) they could provide a matching framework between Web directories, and *ii*) they could be used to assess the semantic structure quality of individual Web directories. In other words, by using them, structures of any two Web directories could be objectively compared and the criteria could point to potential semantic deficiencies in a directory. Information retrieval in Web directories can be executed either through searching or browsing scenarios.

Because of the sheer size of data available on the Web, searching is the dominant information retrieval scenario today. Several performance measures for evaluation of searching scenarios have already been proposed, such as precision, recall, fall-out and F-measure. However, information seeking by browsing scenarios is interesting in reduced information collections like blogs, RSS feeds, social networks [17], but also in individual directory categories. Since information in Web directories may be browsed by intelligent agents as well as human users, the establishment of parameters for objective measurement of Web directory's structure and content is of a significant importance for determining its usability, semantic quality and subsequently other intrinsic characteristics.

We have identified at least three parameters that can be used to objectively assess the semantic quality of a Web directory:

1. Path ratio
2. Maximum revisit
3. Distance decrease progression

All parameters require observation of the browsing pattern of a person or an intelligent agent using the directory. We will assume that the browsing scenario starts at the root category although this is not strictly necessary (nor is often the case in real-world use). The parameters are calculated based on observation of a single Web directory user. Each observation represents one browsing session for a specific resource contained within the directory. After the parameters of individual observations are collected they may be statistically processed and aggregated. This data can then cumulatively represent relevant trends and features in the actions of any number of directory's users.

### 6.1   Path ratio parameter

Path ratio (PR) is calculated as a proportion between the minimum number of categories between the root and the category with the required Web resource, and the number of categories the user traverses while browsing. Therefore, when browsing for a resource $r$ in a Web directory $wd$ the browse $b(r,wd)$ with the length $|b(r,wd)|$ parameter PR is defined as

$$PR(b) = 1 - \frac{\min|b(r,wd)|}{|b(r,wd)|}, PR(b) \in [0,1\rangle \tag{11}$$

The rationale behind this parameter is that in the case of the optimal, or direct, browse $b^*$ the user will achieve the shortest path between the root and the category with the resource browsed for. Therefore path ratio for the optimal search is $PR(b^*)=0$. In a suboptimal, or indirect, browse $b'$ user will traverse at least one category more then $b^*$ and $PR(b')>0$. This is explained in the next figure that illustrates a browsing pattern staring at category $c_1^1$ and ending at $c_9^4$.

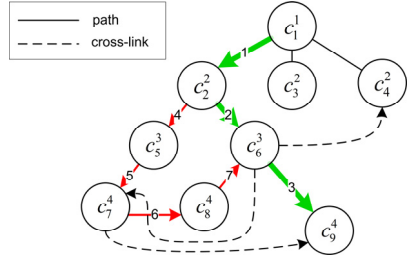

**Fig. 8.** Optimal (direct) and suboptimal (indirect) browse paths in calculating parameter PR.

Browse $b_1$ with the path 1→2→3 is optimal because it traces the shortest and the most direct path between the start and the end category so that $PR(b_1)=0$. While the browse 1→4→5→6→7→3 will also lead to the end resource, it is suboptimal since its length is greater than that of the optimal browse (6 > 3), thus $PR(b_2) = 1/2$.

### 6.2    Maximum revisit parameter

Maximum revisit (MR) or maximum category revisit is a parameter that describes the maximum number of repeated visits to any category while browsing for one resource. Because Web directories are simple rooted graphs with at least one path between any two nodes, there is never a need to visit the same category twice while browsing for a resource. Therefore, MR specifies the level of wandering or loitering in a Web directory's structure while browsing (Fig. 9).

The best possible browse $b$ for a resource $r$ in a Web directory $wd$ has

$$\mathrm{MR}\left(b(r, wd)\right) = 0 \tag{12}$$

indicating no category revisit, where MR(*b*) can be any natural whole number including zero $\mathrm{MR}(b) = 0, 1, 2, ..., n, n+1 \in N \cup \{0\}$.

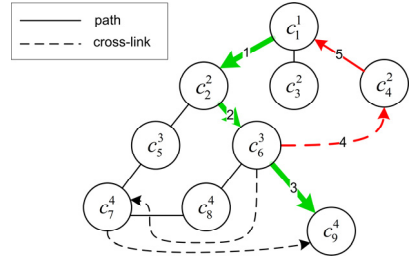

**Fig. 9.** Optimal and suboptimal browsing paths with revisits in calculating parameter MR.

In the Fig 6., browse $b_1$ with the path 1→2→3 starting in $c_1^1$ and finishing with $c_9^4$ has MR($b_1$) = 0. However, due to the configuration of the directory it is possible to needlessly revisit some or even all categories. This is illustrated in the browse $b_2$ with the path 1→2→4→5→1→2→3 which gives MR($b_2$) = 1. Since MR($b_1$) < MR($b_2$) browse $b_1$ is better then $b_2$.

### 6.3   Distance decrease progression parameter

Distance decrease progression (DDP) is an ontology-based parameter. It describes the gradient of semantic convergence toward the resource during one browse. As the user browses categories looking for a particular resource, each category s/he visits should be progressively ontologically closer to the resource. If this is not the case, than either s/he is loitering or the directory does not have the optimal structure. Parameter DPP($s$) can be defined as a series

$$\text{DPP}(b) = \sum_{i=1}^{n-1} dist(c_i, c_T) - dist(c_{i+1}, c_T) \qquad (13)$$

where $c_T$ is the target category containing the resource the user is looking for, $c_i$ is any category being browsed and $n$ is the length, i.e. number of steps, of the browse $b$. It is also necessary to apply a similarity measure $sim : \boldsymbol{C}^2 \to [0,1]$ between the two categories $c_1, c_2 \in \boldsymbol{C}$ and a distance function $dist(c_1, c_2) = 1 / sim(c_1, c_2)$ in [14][15][16]. If the sequence of partial sums $\{s_1, s_2, ..., s_n, s_{n+1}, ...\}$ converges, than the series is also convergent, where

$$s_n = \sum_{k=1}^{m} dist(c_k, c_T) - dist(c_{k+1}, c_T) \qquad (14)$$

The search $b'$ is optimal if DPP($b'$) converges to 0.

All three parameters described here should be used in conjunction with each other in order to cumulatively describe this important design feature of Web directories.

### 6.4   Additional parameters

Except the three parameters explained previously, it is indeed possible to define additional criteria which measure browsing adequacy of a Web directory's structure.

As for an additional parameter, one can measure semantic distance between different pairs of nodes along the browsing path. Instead of only monitoring the distance between the current node and the goal node, it may be prudent to observe change in the semantic distance among the root and the current node. If a directory's structure is truly optimized for browsing, this distance should monotonously grow as the user progressively gets closer to his target resource. Semantic difference between other significant nodes could also be measured and indicate the advance in browsing. These pivotal nodes could be contextually important directory's categories such as the top nodes of individual categories and subcategories, or nodes which have a multitude of cross-links and represent nexuses to other categories. They can also be or structurally important as the right or the left most categories at a certain directory level. Any node that is semantically or structurally unique can be a good point for determining distances to the user's current node and his target node. One can even triangulate between these three points and thus, using geometry, gain further insight in the browsing progress.

Also, it is possible to construct multiple statistical features using any or all previously described parameters and use them as indicators to track the user's resource browsing pattern. The number of features can be increased further by monitoring distances among the current node, the target node, the root and, perhaps, some pivotal node. The idea would be to establish a network within the directory's structure connecting all semantically significant nodes and the user's current node. This would make it possible to dynamically follow the user's actions and how they semantically relate to the goal node.

Finally, with smaller and simpler directories it is entirely possible that they do not have cross-links between categories but only parent-child connections. It would be interesting to compare the efficiency of goal-directed browsing in such structure versus the same structure with additional category cross-links. Intuitively, the idea would be to find the optimum balance between the cross-links and customary parent-child edges. Higher number of cross-links would certainly increase the interconnectivity of the directory's graph facilitating the transits between various categories, but would also make it harder to select the optimum (i.e. shortest) path to the goal category.

### 6.5   Applying the parameters

Node distribution in some Web directories, at a certain level in their structures, does not necessarily have to follow concept semantics partition or this process can be somehow affected and skewed. Examples of this are content division according to date, contributors' names or alphabet, e.g. having node "A" for subnodes with "Apples" content, "B" for "Bananas", "C" for "Citrus", etc. These nodes would have more in common with a concept "Fruit" that with "Alphabet Letter". Subsequently, mutual semantic distance of such nodes would be great and incompatible with the directory's partition. In order to overcome this problem in calculation of the semantic quality parameters one has to simply ignore semantic value of these nodes at a level *l* and directly link nodes in levels *l-1* and *l+1*, i.e. immediately above and below the level *l*. With this monotone semantic difference between nodes can be restored.

Every Web directory should have an easily understandable semantic schema that is reflected in a directory's structure so it becomes self-explanatory which category to browse in order to iteratively and progressively approach the required resource. This issue is closely correlated to the Web usability of directories. However, due to diverse quality of data sources available on the Web it is not easy to construct a directory with an ideal path ratio, maximum revisit and distance decrease progression values. Further planned experiments should provide more information on the everyday applicability of the parameters proposed here.

## 7   Related work

All previous work regarding coupling of Web directories with ontologies and the Semantic Web paradigms has been directed at using Web directories, their data and structure, to extract information from WWW with the goal of document classification and ontology learning. In this paper we presented an exactly opposite approach – using available knowledge to construct a Web directory itself.

The paper by Kavalec [4] which described a mechanism for extraction of information on products and services from the source code of public web pages was especially useful in our work. Papers by Mladenić [5], Li [18] and Brin [19] were also helpful.

Open Directory RDF Dump (http://rdf.dmoz.org/) is an interesting effort because it combines the well-known Web directory Open Directory Project (DMOZ) with Resource Description Framework (RDF) ontology language. DMOZ is the largest human-edited directory found on the Web. Its data is shared by a number of different Web sites and often used by researchers when dealing with knowledge representation in taxonomies. However, the RDF data is read-only and available exclusively off-line in large downloadable packages. It would be much better if the data was accessible in smaller chunks and on demand through a dedicated Semantic Web service. Also, any usage of ontologies (even lightweight ones stored in RDF only) presents a new set of problems such as ontology mapping, alignment, discovery, etc. The DMOZ administrators should do more regarding integration of this valuable data with the Semantic Web vision.

We would particularly like to emphasize the work by a research group at University of Zagreb which introduced ontologies in the search mechanism of the Croatian Web directory, and thus successfully resolved problems of low recall, high recall and low precision and vocabulary mismatch [20]. The Croatian Web directory

(http://www.hr/) [13] was founded in February 1994 and its purpose has been to promote and maintain the network information services through the "national WWW homepage" and enable easy navigation in Croatian cyberspace using hierarchically and thematically organized directory of WWW services. At the moment of writing the directory contains over 25,000 Web resources listed in more than 750 categories.

## 8   Acknowledgment

This work has been supported by the Croatian Ministry of Science, Education and Sports through research projects "Content Delivery and Mobility of Users and Services in New Generation Networks" and "Adaptive Control of Scenarios in VR Therapy of PTSD", grant numbers 036-0362027-1639 and 036-0000000-2029, respectively. The authors would also like to cordially thank the reviewers for their most helpful suggestions which contributed to the improvement of the paper.

## 9   Conclusion

Web directories are a commonplace method for structuring various semantically heterogeneous resources. The form of simple rooted graphs is well-suited for many uses in information storage, representation and retrieval in Web and desktop environments, such as is-a hierarchies, taxonomies, directory trees, bookmarks, menus, and even emotionally annotated databases [21]. Also, the aspect of social collaboration is very important since networking and the Web enable instant publication and usage of data – ideally within groups of trusted users with shared areas of interest. All this facets emphasize the importance of successful construction, management, information extraction and reuse of information stored within Web directories' structure.

The first goal of this paper was to devise a generic method for automatic construction and maintenance of Web directories content and structure. The second goal was to propose a set of objective criteria that can be used for appraisal of directories structure utility in browsing-based information retrieval. The first two of these parameters (path ratio and maximum revisit) are based on heuristics while the third (distance decrease progression) requires introduction of ontologies in description of knowledge contained in the Web directory, which is possible only if Web directories are placed in the context of the Semantic Web vision.

We recommend caution in using publicly available Web directories to learn new ontologies. Structures of Web directories are often biased and greatly influenced by contributors and the order in which they added resources to directories. Also, the maintenance of a large directory is an overwhelming task prone to errors. Therefore, it may be more advisable to construct ontologies from smaller directories or from directories with rigid administrative policies. Small directories are more numerous than the larger but they will offer less information and in more specialized and segmented areas.

In the future work we would like to expand the initial system and build a hard general ontology which would efficiently encompass smaller ontologies of individual categories and provide a unitary base for ontology matching throughout the Web directory. Furthermore, we would like to test the upgraded system in real-life situations and use it regularly as a decision support system in maintenance of a large Web directory. In the near future we are planning to validate the system and evaluate its features by implementing it within the Croatian Web directory and its domain "Tourism" as a suitable test category. The semantic quality measures would be used as control parameters in an iterative process of constructing and perfecting the Web directory's structure. In the first phase of this experiment we are set to develop a semi-automatic system which only proposes a choice of optimal recommendations without taking explicit action by itself. This would give us an opportunity to fine tune the algorithms and ascertain their practical usefulness.

Also, in the mid-term future we see an opportunity to switch from tag cloud based resource annotation used in many popular Web 2.0 sites such as YouTube, Flickr, IMDB, del.icio.us, Amazon, etc. toward lightweight ontologies. In this case the ontology instances could be browsed in a manner similar to the use of Web directories and the semantic quality measures, as well as the ontology-based directory construction processes, could be used to improve or facilitate the information extraction in such Web sites.